\documentstyle[prl,aps,multicol,graphics]{revtex}

\begin{document}
\title{Exclusion of quantum coherence as the origin of the 2D metallic
state in high-mobility silicon inversion layers}
\author{G.~Brunthaler$^{(a)}$, A.~Prinz$^{(a)}$, G.~Bauer$^{(a)}$,
and V.~M.~Pudalov$^{(a,b)}$}
\address{$^{(a)}$ Institut f\"{u}r Halbleiterphysik, Johannes Kepler %
Universit\"{a}t, A-4040 Linz, Austria\\ %
$^{(b)} $P.\ N.\ Lebedev Physics Institute of the Russian Academy %
of Sciences, Moscow 117924, Russia }
\date{\today}
\maketitle

\begin{abstract}
The temperature and density dependence of the phase coherence time
$\tau_\varphi$ in high-mobility silicon inversion layers was
determined from the magnetoresistivity due to weak localization.
The upper temperature limit for single-electron quantum
interference effects was delineated by comparing $\tau_\varphi$
with the momentum relaxation time $\tau$.
A comparison between the density dependence of the borders for
quantum interference effects and the strong resistivity drop
reveals that theses effects are not related to each other. As the
strong resistivity drop occurs in the Drude regime, the apparent
metallic behavior can not be caused by quantum coherent effects.
\end{abstract}


\begin{multicols}{2}

The apparent ``metallic'' state in two dimensions (2D)
\cite{Krav94} has attracted much attention as it seems to
contradict the one parameter scaling theory of Abrahams et
al.\,\cite{Abra79}. Following the confirmation of the ``metallic''
behavior in several material systems, the question was raised
whether the ``metallic'' state constitutes a new quantum
mechanical ground state, or if the resistivity drop towards lower
temperature is based on semiclassical (i.e., noncoherent) effects
(see \cite{Abrahams2000RMP} and references therein).

We answer this question for Si-metal oxide semiconductor (MOS)
structures, by excluding quantum interference (QI) effects as the
origin of the ``metallic'' state. This is achieved by determining
the phase coherence time $\tau_\varphi$ from the weak localization
(WL) behavior and comparing it with the momentum relaxation time
$\tau$ at different temperatures $T$ and densities $n$. For
$\tau_\varphi
> \tau$ (low-$T$ regime), single-electron quantum
interference effects occur, whereas for $\tau_\varphi < \tau$
(high-$T$) they do not, as the coherence time is too short to
allow electrons a coherent return to their origin. By comparing
the phase coherent regime with the ``metallic'' regime, we find
that they are not correlated with each other and that ``metallic''
behavior exists even without phase coherence. In addition, we
dertermine the temperature where $k_B T = \hbar / \tau$, which
marks the threshold for coherent electron-electron ($e-e$)
interaction effects. Again, no correlation with the ``metallic''
regime is found.

The $T$ dependence of the phase coherence was already investigated
in the early 80's in Si-MOS structures (see
\cite{Davies83,Kawaji86}). But due to the lack of the ``metallic''
state in these low-mobility samples, no appropriate conclusion
could be drawn. In recent studies on high-mobility samples with
``metallic'' behavior, it was shown that the WL has only small
effects on $\rho$ for GaAs/AlGaAs \cite{SimmonsPRL00} and Si/SiGe
\cite{Senz00}. Also for Si-MOS structures in the low $\rho$ (high
$n$) regime, the WL contribution is small and it was shown that
spin-orbit coupling is not visible for $\tau_\varphi$ up to
100\,ps \cite{Bru99}. But so far, the borders for phase coherence
were not determined systematically on a sample with strong
``metallic'' behavior in order to decide among the possible
underlying mechanisms.

Our investigations were performed on two high-mobility Si-MOS
samples Si-15 and Si-43 with peak mobilities of $\mu = 31\,000$
and 20\,000\,cm$^2$/Vs, respectively. Resistivity and Hall
measurements were performed with a four terminal ac-technique at a
frequency of 17.17\,Hz.

Figure 1 shows the magnetoresistivity $\rho(B)$ of sample Si-15
for temperatures between 280\,mK and 1.59\,K at $n$ = $3.7 \times
10^{11}$\,cm$^{-2}$. The peak in $\rho(B)$ can be fitted by the
conductivity corrections arising from single-electron coherent
backscattering (weak localization) according to \cite{WL-theory}
\begin{equation}
\Delta\sigma_{xx} = - \frac{\alpha g_\nu e^2}{2\pi^2\hbar}
 \left[\Psi\left(\frac{1}{2}+\frac{a}{\tau}\right)
       -\Psi\left(\frac{1}{2}+\frac{a}{\tau_{\varphi}}\right)\right],
\label{eq:WL}
\end{equation}
where $\Psi$ is the Digamma function;  $a = \hbar/4eBD$ with $B$
the applied perpendicular magnetic field and $D$ the diffusion
coefficient. The values for $D$ and $\tau$ were deduced from our
Hall and $\rho$ measurements, assuming at first that the linear
Drude regime holds (this restriction will be omitted later on).
The prefactor $g_{\nu}=2$ describes the valley degeneracy for
(100) Si-MOS and $\alpha$ depends on the ratio of intra-valley to
inter-valley scattering rates and should lie between 0.5 and 1
\cite{Fuku81}. We found values between 0.55 and 0.8, i.e., inside
the expected range. The solid lines in Fig.~1 are least square
fits of Eq.~(1) to the data.

For sample Si-15 the WL-peak was followed down to small $n$ of
about $1.5 \times 10^{11}$\,cm$^{-2}$, near the
``metal-insulator'' transition at $n_c \approx 0.8 \times
10^{11}$\,cm$^{-2}$. At still smaller $n$, the rising contact
resistance decreases the signal-to-noise ratio so that the WL
cannot be evaluated any more. Sample Si-43 showed a similar
behavior.

The temperature dependence of the phase relaxation time
$\tau_\varphi$ and of the momentum relaxation time $\tau$ is
depicted for sample Si-15 in Fig.~2 for four different $n$ between
$1.93 \times 10^{11}$ and $1.03 \times 10^{12}$\,cm$^{-2}$. At low
$T$, $\tau_\varphi$ exceeds $\tau$ by nearly up to two orders of
magnitude. But since $\tau_\varphi$ strongly decays with
temperature, $\tau_\varphi$ and $\tau$ cross each other at higher
$T$. When $\tau_\varphi < \tau$ the electrons can not return to
their origin within the phase coherence time and no
single-electron QI is possible.  Thus the temperature $T_q$ at
which $\tau_\varphi = \tau$ defines an upper limit for
single-electron quantum interference.  This definition is in
agreement with our experimental observation that the WL-peak
vanishes just below $T_q$.  A very similar behavior was found in
sample Si-43, where we traced the WL-peak for $n$ between $5.4
\times 10^{11}$ and $3.5 \times 10^{12}$\,cm$^{-2}$ and up to a
maximum $T$ of 10.7\,K at high $n$.

\begin{figure}
\begin{center}
\resizebox{0.80\linewidth}{!}{\includegraphics{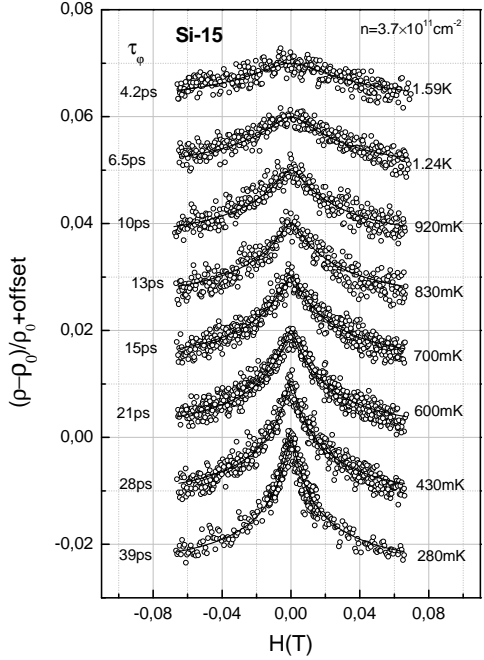}}
\begin{minipage}{8.5cm}
\vspace{0.2cm} \caption{Change of resistivity versus perpendicular
magnetic field for sample Si-15 for eight temperatures at $n = 3.7
\times 10^{11}$\,cm$^{-2}$. The solid lines are fits to weak
localization according to Eq.~(1). }
\end{minipage}
\end{center}
\label{fig:Si15rho(B)}
\end{figure}
\vspace{-0.3cm}

For the determination of $\tau_\varphi$, we used $D$ and $\tau$ as
calculated from the Hall and $\rho$ measurements, assuming the
Drude regime to hold. This is \emph{a priori} not justified,
because $\tau$ might remain at it's ``high''-$T$ value and the
strong changes in $\rho(T)$ may originate from quantum
interference effects. Although we know, that weak localization
gives only small contributions to $\rho(T)$, there might be
additional corrections based on quantum interference. In order to
test the above results, we have thus evaluated additionally $T_q$
and $\tau$ directly from the WL-peak without the use of $D$, i.e.,
without assuming the Drude regime to hold. The phase coherence
length $\ell_\varphi =\sqrt{D \tau_{\varphi}}$ follows directly
from a fit of the curvature and width of $\rho(B)$ near $B = 0$,
independent of $D$ and $\tau$.

The mean free path $\ell$ is then obtained from the height of the
WL-peak which is proportional to $\ln(\ell_\varphi / \ell)$. From
$\ell_\varphi(T)$ and $\ell(T)$, we find the crossing point $T_q$
at which $\ell_\varphi = \ell$. For $n = 3.7 \times
10^{11}$\,cm$^{-2}$ we find, e.g., 2.1\,K, which is practically
the same as obtained from the crossing of $\tau_\varphi$ and
$\tau$, i.e., 2.2\,K (see Fig.~2). This shows that the temperature
limit $T_q$ does not depend on the assumption that the Drude
regime is effective and can be deduced solely from the decrease of
the WL peak towards higher $T$. Knowing $\ell$, we can deduce a
value for $\tau$ from $\tau = \ell / v_F$, with $v_F$ being the
Fermi velocity. E.g., for $n = 3.7 \times 10^{11}$\,cm$^{-2}$, we
obtain a value between 3 and 5\,ps at $T = 0.3$\,K, depending on
the prefactor $\alpha$. This range for $\tau$ is in good agreement
with the Drude value of $\tau = 2.85$\,ps as deduced from
$\rho(T)$ at low $T$ ($< 2$\,K) and far away from the ``high''-$T$
value of 0.45\,ps (at $T \approx 40$\,K in Fig.\ 2). An estimate
of $\tau$ exclusively from the weak localization thus leads to the
same value as it is obtained assuming the Drude relation $\sigma =
ne^2\tau / m^*$. This consistency directly proves that the
metallic state indeed obeys Drude behavior.

\begin{figure}
\begin{center}
\resizebox{0.75\linewidth}{!}{\includegraphics{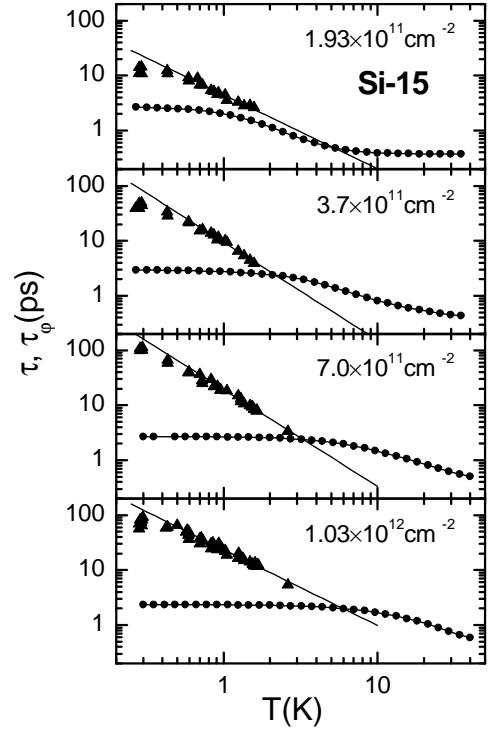}}
\begin{minipage}{8.5cm}
\vspace{0.2cm} \caption{Experimental temperature dependence of
phase coherence time $\tau_\varphi$ (triangles) and momentum
relaxation time $\tau$ (circles connected by line) for sample
Si-15 at $n = 1.93$, 3.7, 7.0 and $10.3 \times
10^{11}$\,cm$^{-2}$. The thin solid line represents a $T^{-p}$
least mean square fit to the data points. }
\end{minipage}
\end{center}
\label{fig:Si15tau(T)}
\end{figure}
\vspace{-0.3cm}

According to theory, the dependence of $\tau_\varphi$ on
conductance and temperature for inelastic $e-e$ scattering in the
limit of small momentum transfer (low $T$) can be described by
$\tau_\varphi = \hbar g / k_B T \ln(g/2)$, where g is the
dimensionless conductance in units of $e^2/h$ \cite{interaction}.
This relation is strictly valid only for $g \gg 1$.  For the case
that $g$ becomes of the order of 1, the term $g/\ln(g/2)$ should
be substituted by something of the order of unity
\cite{Alts-priv}. By fitting $g/\ln(g/2)$ at $10 \leq g \leq 100$
we obtained the second order polynomial $f(g) = 3.78 + 0.253 g -
0.00036 g^2$ and used this in the above expression.

We find that the calculated $\tau_\varphi(T)$ has a smaller slope
than the experimentally determined one. In addition, the
calculated values are nearly a factor of 10 larger than the
experimentally determined ones, even for $g > 10$. The
experimental data can be fitted much better with a $T^{-p}$ law
(solid lines in Fig.~2). It seems that the electron system is not
in the pure inelastic $e-e$ scattering limit with small momentum
transfer. The $T^{-p}$ dependence with $p > 1$ points to a
$T^{-2}$ contribution resulting from $e-e$ scattering processes
with large momentum transfer (pure metal case) \cite{Fukuyama83}.
We find $p$ between 1.1 and 1.7, depending on $n$, similar to
earlier values in Si-MOS (see \cite{Kawaji86,Davies83} and
references therein). For our purpose, it is important that the
intersection of the extrapolated $T^{-p}$ dependence with the
momentum relaxation time $\tau$ gives a well defined value for the
temperature limit $T_q$ for single-electron QI (see Fig.~2).

\begin{figure}
\begin{center}
\resizebox{0.95\linewidth}{!}{\includegraphics{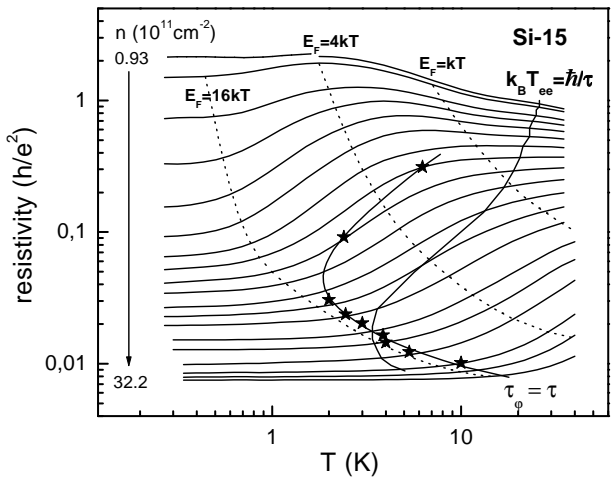}}
\begin{minipage}{8.5cm}
\vspace{0.2cm} \caption{Temperature thresholds in the $\rho$
versus $T$ plane for Si-15. The asterisks mark the threshold $T_q$
for single-electron quantum interference, and the $k_B T_{ee} =
\hbar / \tau$ line indicates the threshold for quantum
interference effects due to $e-e$ interaction. Dashed lines mark
$E_F/i = k_B T$, with $i = 1, 4,$ and 16. Experimental $\rho$ vs.\
$T$ curves are shown for $n = 0.928$, 0.970, 1.09, 1.18, 1.30,
1.45, 1.64, 1.89, 2.22, 2.64, 3.18, 3.88, 4.79, 6.30, 7.95, 10.2,
15.7, 21.2, and $32.2 \times 10^{11}$\,cm$^{-2}$. }
\end{minipage}
\end{center}
\label{fig:Tlimits}
\end{figure}
\vspace{-0.3cm}

Figure 3 shows the main feature of the ``metallic'' behavior in
Si-MOS structures, i.e., the strong drop in $\rho(T)$. The
temperature limit $T_q$ for single-electron quantum effects is
marked in Fig.~3 by asterisks, connected by a curve to guide the
eye. Depending on density, $T_q$ lies between 2 and 10\,K. At high
density, $T_q$ decreases with decreasing $n$, but when $n$ drops
further $T_q$ returns to higher values. This retrograde behavior
is the consequence of the strong decrease of $\tau$ in the low $n$
range, which moves the crossing of $\tau_\varphi$ and $\tau$ to
higher $T$ as can be seen in Fig.~2 for $n = 1.93 \times
10^{11}$\,cm$^{-2}$.

An important conclusion can be drawn by comparing the overall $T$
dependencies in Fig.~3. The strong drop in $\rho(T)$ occurs for
low $n$ between 0.6 and 2\,K and shifts with increasing $n$ to
much higher $T$. For the highest $n$ of $3.2 \times
10^{12}$\,cm$^{-2}$ the decrease in $\rho$ lies completely above
10\,K. This behavior does not correspond to the observed
dependence of the QI threshold $T_q$ which first decreases and
then increases with increasing $n$. For $n > 2.2 \times
10^{11}$\,cm$^{-2}$ the strong drop in $\rho(T)$ takes place
entirely above the single-electron QI limit $T_q$ and thus must be
caused by other effects.

We further indicate the temperature $T_{ee}$ in Fig.~3, which is
defined by $\hbar / \tau = k_B T_{ee}$ and gives an upper limit
for the occurrence of QI corrections due to $e-e$ interaction
\cite{interaction}. Again, the run of the temperature limit
$T_{ee}$ does not coincide at all with the density and temperature
behavior of the strong change in $\rho(T)$. We find again that for
large $n$ the strong drop in resistivity is above the quantum
correction border $T_{ee}$ and thus the ``metallic'' state can not
be caused by $e-e$ induced QI effects either.

As there is no room for QI effects at high densities, the strong
$\rho(T)$ drop has to be generated by semi-classical effects. But
even at smaller densities, where the resistivity drop is quite
similar, it is not expected that its origin is suddenly changing
from non-QI to QI. A strong influence of semi-classical effects
should extend even to much lower densities.

On the other hand, negative magnetoresistance due to weak
localization was found for $T < T_q$ at all densities,
demonstrating the existence of small single-electron QI effects at
low $T$. This behavior is in contradiction to the suggested
superconductivity of the ``metallic'' state \cite{Phillips98},
where no single-electron QI is expected.

We also indicate the relations $E_F = k_B T$, $E_F = 4k_B T$, and
$E_F = 16k_B T$ in Fig.~3, which are related to electron
degeneracy. It is worth to note that the bulk of the resistivity
changes takes place along the $E_F = 4k_B T$ line for a very large
density range which in itself favors a semi-classical explanation
for the strong resistivity drop. In addition, we find that the
equality $E_F = 16k_B T$ is relatively close to the low-$T$
saturation of $\rho(T)$, although small changes persist to even
lower $T$.

As the strong $\rho(T)$ drop at high densities is caused by
semi-classical effects, we discuss several mechanisms. For
p-Si/SiGe samples, with small changes of about 10\% in $\rho(T)$,
it has been shown that the ``metallic'' behavior can be explained
by temperature dependent screening effects for impurity scattering
\cite{SenzCondmat00}. But as in the Si-MOS system, the observed
changes in $\rho(T)$ amount up to a factor 10, the question is how
large the contribution of screening can be. Das Sarma and Hwang
have calculated numerically that indeed changes by an order of
magnitude may occur in $\rho(T)$ \cite{DasSarmaPRL99}. Very recent
calculations of Gold \cite{GoldJETPL2000} indicate that
exchange/correlation and multiple scattering effects dominate the
screening behavior for small $n$ and that the large ratio of up to
10 in $\rho(B_c)/\rho(B=0)$ in parallel magnetic field can also be
explained, where $B_c$ is the field for complete spin polarization
\cite{Kawaji}. In agreement with the expected screening behavior,
a linear $T$ dependence in $\rho(T)$ has been observed in high
mobility Si-MOS samples in the intermediate $T$ range at low $n$
\cite{AltshulerCondmat0008005}.

The scattering of electrons at charged hole traps in the oxide
layer of Si-MOS is also able to explain a strong $T$ and $B$
dependence of $\rho$ in the frame of semi-classical effects
\cite{AltsPRL99}. The filling of hole traps and thus the
efficiency of scattering depends strongly on the Fermi energy
$E_F$ which depends on $n$ and $T$. The low-$T$ saturation of
$\rho$ in the ```metallic'' regime is in this model an interplay
between the neutralization of the charged traps (for $E_F$ above
trap energy) and the $T$-independent scattering at low
temperatures by surface roughness and residual impurities. In a
recent paper, it was shown that the density of defect states on
the Si/SiO$_2$ interface has a large influence on the $\rho(T)$
behavior of the system \cite{Safonov}.

As a third mechanism also band splitting may give rise to strong
variations in $\rho(T)$. In p-GaAs/AlGaAs, temperature dependent
interband scattering \cite{Yaish99} and anomalous
magneto-oscillations \cite{WinklerPRL00} have been observed in the
``metallic'' state. These effects are related to the strong
spin-orbit interaction together with the inversion asymmetric
confinement potential, which induces a splitting of the upper
valence band. But spin-orbit interaction is very weak in Si-based
structures \cite{WinklerCondmat2000} and does not lead to large
changes in $\rho(T)$ in our system. Recently, a small valley
splitting at magnetic fields $B \rightarrow 0$ was extracted from
precise Shubnikov-de Haas investigations
\cite{PudalovCondmat2001}. It was found that the mobilities in the
two valleys are very similar and thus cannot cause large changes
in $\rho(T)$ either.

For the large range of carrier densities investigated in our
Si-MOS structures, we attribute the strong ``metallic'' decay of
$\rho(T)$ both to carrier screening and impurity scattering
effects. The metallic state is observed over nearly two orders of
magnitude in electron density and thus the relative strength of
these different mechanisms will vary. Valley splitting is ruled
out as the mechanism causing the large drop in $\rho(T)$.

In conclusion, we have answered the fundamental question about the
origin of the``metallic'' state by showing that the strong
resistivity drop exists without the presence of quantum effects.
The borders for phase coherence were deduced from the temperature
dependence of the weak localization at different densities.   For
densities above $2.2 \times 10^{11}$\,cm$^{-2}$, the decrease of
the resistivity into the ``metallic'' regime takes place in the
absence of phase coherence.  Also disorder induced quantum
interference effects due to electron-electron interaction cannot
be the origin of the ``metallic'' state as the boundary $\hbar /
\tau = k_B T_{ee}$ is not related with the resistivity drop
either.  Thus semi-classical effects are responsible for the low
resistivity state over a very large carrier density range in the
Si-MOS system, where the temperature dependence of the resistivity
is the strongest of all 2D systems.

We thank B.\,L.~Altshuler and A.~Gold for stimulating discussions.
The work was supported by the Austrian Science Fund (FWF) Project
P13439, INTAS (99-1070), RFBR, NATO (PST.CLG.976208, NSF
(0077825), Programs \lq Physics of solid state nanostructures',
\lq Statistical physics' and \lq Integration'

\end{multicols}
\end{document}